\documentclass[runningheads]{llncs}

\usepackage{graphicx}

\usepackage{floatrow}
\newfloatcommand{capbtabbox}{table}[][\FBwidth]
\usepackage{appendix}
\usepackage{amsmath}

\usepackage{comment}

\begin{document}
\title{High Recall, Small Data: The Challenges of Within-System Evaluation in a Live Legal Search System\thanks{The authors would like to thank the respondents for their participation in this research. The authors would also like to thank Legal Intelligence, in particular T.E. de Greef and P. van Boxtel, for their cooperation. \\\\
\textbf{Statements and Declarations}\\
{At the time of writing G. Wiggers was affiliated with Leiden University as a PhD candidate, and with Wolters Kluwer (Legal Intelligence) as business analyst. At the time of research R. van der Burg was affiliated with Radboud University as a master's student, and wrote his thesis in cooperation with Wolters Kluwer (Legal Intelligence). He is currently affiliated to Wolters Kluwer as Senior Product Software Engineer.\\}}}
\titlerunning{High Recall, Small Data}
\author{Gineke Wiggers\inst{1,2}\orcidID{0000-0002-1513-2212} \and Suzan Verberne\inst{1}\orcidID{0000-0002-9609-9505} \and Arjen de Vries\inst{3}\orcidID{0000-0002-2888-4202} \and Roel van der Burg\inst{2,3}}
\authorrunning{G. Wiggers et al.}
\institute{Leiden University, Leiden, The Netherlands\\
\email{\{g.wiggers@law., s.verberne@liacs.\}leidenuniv.nl}\\
\and Wolters Kluwer (Legal Intelligence), Alphen aan den Rijn, The Netherlands\\
\email{roel.van.der.burg@wolterskluwer.com}\\
\and Radboud University, Nijmegen, The Netherlands\\
\email{a.devries@cs.ru.nl}
}

\maketitle              % typeset the header of the contribution

\begin{abstract}
This paper illustrates some challenges of common ranking evaluation methods for legal information retrieval (IR). 
We show these challenges with log data from a live legal search system and two user studies.
We provide an overview of aspects of legal IR, and the implications of these aspects for the expected challenges of common evaluation methods: test collections based on explicit and implicit feedback, user surveys, and A/B testing. Next, we illustrate the challenges of common evaluation methods using data from a live, commercial, legal search engine. 
We specifically focus on methods for monitoring the effectiveness of (continuous) changes to document ranking by a single IR system over time.
We show how the combination of characteristics in legal IR systems and limited user data can lead to challenges that cause the common evaluation methods discussed to be sub-optimal.
In our future work we will therefore focus on less common evaluation methods, such as cost-based evaluation models.

\end{abstract}

\section{Introduction}

\par In the legal domain, the amount of information available digitally is increasing rapidly. Legal scholars and professionals have to navigate this information to find the case law and articles relevant for them. They often do this under the time pressure of having to account for every minute spend on a case. A study by LexisNexis showed that attorneys spend approximately 15 hours in a week seeking case law \cite{Lastres}. Legal information retrieval (IR) systems exist to help legal professionals navigate this information overload to find relevant information in the most efficient way. In order to do this, legal IR systems are continuously improving their retrieval and ranking algorithms. Evaluation of these systems is important from a commercial and academic point of view; however, in practice this is not always conducted in a consistent manner.

\par That evaluation of legal IR is not always conducted in a consistent manner was shown by Conrad and Zeleznikow in their work on the use of evaluation methods in articles on legal IR in the ICAIL proceedings \cite{Conrad2} and the journal Artificial Intelligence and Law \cite{Conrad}. They find that ``there may remain some cause for concern insofar as a scientific research community that champions Artificial Intelligence for the benefit of the legal domain may still have as many as a fifth of its empirical conference works presenting no performance evaluation at all." \cite[p. 185]{Conrad} Aside from this one fifth missing evaluation at all, their results show that 46\% of the papers use gold data created by domain experts as evaluation method and a further 22\% use manual assessment by grad students or research assistants. Conrad and Zeleznikow argue that if the research community in AI and law wishes to remain relevant to legal practitioners, they have to develop methods to show the value of their work \cite{Conrad}. {This would mean including evaluation in every paper, and perhaps moving towards evaluation involving end users.}

\par In this paper we show that evaluation of legal IR systems is not only lacking for certain research settings, but that the challenges causing these missing evaluations also exist for live legal IR systems. We describe evaluation challenges and limitations based on the literature about legal IR and illustrate why the common evaluation approaches do not work for live professional search systems. We do so using data from a live legal IR system and two exploratory user studies. We focus on within-system evaluation of changes in ranking algorithms.
This applies to situations where a change to the algorithm is made that affects the ranking of the documents but not the number of documents retrieved to allow scholars and developers to assess the effect of the change in the ranking algorithm. We address the following research questions:
\begin{enumerate}
    \item What are the characteristics of legal IR that influence the choice of ranking evaluation methods and metrics?
    \item What are the challenges of common evaluation methods and metrics for evaluating ranking changes in live professional IR systems?
\end{enumerate}

\par The data for our work is provided by Legal Intelligence, one of the largest legal content aggregators and legal IR systems in the Netherlands.

\par The contribution of this paper is to demonstrate why common ranking evaluation methods are difficult to apply to live professional search systems. We do this by (1) providing insight in the characteristics of legal IR in practice that make the task different from common ranking evaluation tasks; 
(2) describing the challenges in applying common evaluation methods to be expected based on these characteristics; and (3) illustrating these challenges using data from a live legal search engine.

\par Legal IR is often compared to other types of professional search with high recall requirements. However, it is useful to remember that the medical domain has English as an overarching language, with platforms like PubMed to cater to a worldwide audience. Similarly patent retrieval focuses on English, German and French as overarching working languages. But most legal professionals will limit their search to information from their own jurisdiction (country) and language. This makes legal IR distinct from other high recall situations.

\par To define which evaluation methods are common, we based ourselves on the classic textbook from Manning et al. \cite{Manning}. We assess the following evaluation methods for our example: (a) a test collection based on information needs and relevance judgments by domain experts, (b) a test collection based on implicit feedback from clickthrough/log analysis, (c) user satisfaction studies (in particular surveys), and (d) A/B testing.

\par In Section \ref{LegalIR} we conduct a literature analysis to answer research question 1. In Section \ref{Expected challenges} we discuss expected challenges in the application of common evaluation methods and metrics for live professional IR systems. In Section \ref{Found challenges} we illustrate, using data from our legal search engine, the challenges in applying these methods. Based on the information from the literature analysis and the data we will conclude in Section \ref{Conclusion} by answering research question 2.

\section{Legal IR}\label{LegalIR}
To understand why common ranking evaluation methods are difficult to apply to legal IR systems, we need to have a clear picture of the characteristics of these systems and their users. This section starts with the description of the characteristics of legal IR, its users and its documents, contrasting its properties with these of Web search where possible. It also relates legal IR to professional IR in general, to further specify the characteristics of legal IR.

\subsection{The User}\label{Users}
The classical image of a legal professional is a lawyer who (1) works under high time pressure and (2) cannot afford to miss information that might be relevant in court. The time pressure for lawyers (and other legal professionals) often stems from the billing system, where every hour or even minute dedicated to a case has to be accounted for. This is often tracked using specific software.\footnote{E.g. Nederlandse Orde van Advocaten: Modelkantoorhandboek. The Hague (2020) Available at: \url{https://www.advocatenorde.nl/kantoorhandboek}} %As one such software provider states: ``Time is money and every unrecorded minute is lost revenue''.\footnote{https://www.bill4time.com/time-tracking-software, accessed October 2020}

\par At the same time, legal professionals cannot afford to miss any important information. Their professional reputation would be damaged if the opposing party has information they have missed. Konstan et al. \cite{Konstan} analyzed the cost-benefit values for different user groups, and show that for legal users, missing an item that turns out to be valuable has a very high negative impact. In contrast, false positives (reading an irrelevant article) have a medium negative impact, and correct negatives (correctly removing articles from the results list) have a low/medium positive impact. This is in line with the conclusion by Bock \cite{Bock} that the main focus in legal IR should lie on high recall.\footnote{See also Mart \cite{Mart}.} Manning et al. \cite[p. 156]{Manning} even go as far as to say that paralegals will tolerate fairly low precision results to obtain this high recall.

\par Geist observes in \cite{Geist} that although high recall is in theory preferred, the reality of the time pressure that all legal professionals perform under means that precision is required. He calls it the `completeness ideal' and the `research reality'\footnote{`Vollständigkeit(sideal) und Recherche-Realität' \cite[p.158]{Geist}, translation by author.}: ``Simply put, it is in a legal dispute first of all important to know more than the opposing lawyer(s) and not to fulfill abstract ideals of completeness''. %\footnote{'Vereinfacht gesagt, ist es in einer gerichtlichen Auseinandersetzung zumindest zunächst nur entscheidend, mehr zu wissen, als die gegnerischen Anwält/innen und nicht abstrakte Vollständigkeitsideale zu erfüllen' \cite[p. 160]{Geist}, translation by author.} 
\par The `completeness ideal' suggests that legal professionals do not stop their research until they have achieved full recall. But the `research reality' suggests that there is a point where the legal professional is `sure enough' and will stop. Where this stopping point depends on the user (e.g. a novice versus a senior lawyer, or a general practice lawyer versus a highly specialised lawyer) and the case at hand. Geist \cite{Geist} argues that only a good relevance ranking can provide users with both high recall (completeness) and high precision (most relevant results first).

\par A secondary effect of the time pressure of legal professionals is that the gathering of explicit feedback (asking users or judges to evaluate search results) is prohibitively expensive for developers of legal IR systems. This leads scholars and developers to use feedback from graduate students \cite{Conrad}\cite{Conrad2}.
%Implicit feedback, which is collected in the course of normal search activities by monitoring the actions of the users \cite{Kelly}, is therefore preferred, as this can be collected at minimum expense and unobtrusively.

\par A practise shown very often by legal professionals \cite{makri2008} and much less in Web search \cite{Kellar}\cite{Teevan} is
\textit{updating}. Updating behaviour refers to gaining understanding about the current importance or status of a particular document \cite{makri2008}. It could be regarded as a type of known-item retrieval: the user is aware of the existence of a document, or a state of a document, and needs to know if their knowledge is still up-to-date. An example of this is monitoring amendments to laws to verify if something is still the accepted interpretation of the law. This updating behaviour is mostly done in a direct way by querying for the particular document or indirectly by means of an automatic citator service \cite{makri2008}. 

\par Van der Burg \cite{VanderBurg} found that of all queries investigated, 25\% is inferred, or assumed known-item search and 75\% are other searches. As they point out this frequency of known-item searches lies close to the 20\% navigational queries found by Broder \cite{Broder}. Van der Burg describes that the queries in the assumed known-item set are on average shorter than those in the remainder set, and that the clicks related to the assumed known-item set are more often on the highest ranked documents \cite{VanderBurg}.

\par Another characteristic of legal professionals is that they wish to have control in their search \cite{Russel-Rose}. Mart \cite{Mart} describes the ranking algorithms of two leading American legal IR systems, Westlaw and Lexis. She explains that companies treat their ranking algorithms as trade secrets, and are therefore reluctant to discuss them in detail, but based on the information she gathered from various sources, it appears that Westlaw considers ``...commercial user document interaction history" \cite[p. 400]{Mart}\cite{Peoples} in their ranking, something that is common in Web search. Lexis on the other hand states: ``This is not a popularity algorithm! Our algorithms provide you with more control over your research..." \footnote{
LexisNexisLawSchools: Understanding the technology and search algorithm behind Lexis Advance. Retrieved at \url{https://www.youtube.com/watch?v=bxJzfYLwXYQ\&feature=youtu.be}} This need for control makes the user requirements for legal IR systems different from those of Web search engines like Google.

\subsection{The IR Systems}
What most legal IR systems have in common, with the exception of a small number of commercial IR systems, is that they limit themselves to one jurisdiction. This limited scope distinguishes legal IR from Web search, but also from other types of professional search. When looking in more detail, legal IR systems can be divided into two broad groups, based on their owners: (1) governments and (2) publishers \cite{Geist}.
\par Governments, in their role as legislative and judiciary branch \cite{Montesquieu}, create laws and case law. These are often published on government websites with an IR system build into it.\footnote{E.g. \url{https://www.govinfo.gov/app/collection/STATUTE} for US Statutes at Large and \url{https://www.loc.gov/collections/united-states-reports/} for selected US Reports, \url{https://www.gesetze-im-internet.de/index.html} for German laws and \url{https://www.bundesverfassungsgericht.de/DE/Homepage/homepage\_node.html} for case law from the German Bundesverfassungsgericht, \url{https://www.ris.bka.gv.at/} for Austrian law and case law, and \url{https://wetten.overheid.nl/zoeken} and \url{https://www.rechtspraak.nl/} for Dutch law and case law.} These systems are often limited to one information type, either law or case law, and in federal government structures often further delimited to federal law/case law or state law.
\par Publishers create commercial legal IR systems to make their publications more accessible to legal professionals on subscription basis.\footnote{Westlaw, an American legal IR system active in many countries is owned by ThomsonReuters, see \url{www.westlaw.com}. LexisNexis, another US based system operating in many countries is owned by the RELX group, formerly known as Reed Elsevier, see \url{www.lexisnexis.com}.
In Austria \cite{Geist}, there is RDB owned by publisher Manz (\url{www.rdb.at}), LexisNexis Austria
(\url{www.lexisnexis.at}), and Linde Digital owned by Linde Publishers (\url{https://www.lindedigital.at/}). Exception to the rule appears to be RIDA created and maintained by prof. Jahnel, see \url{http://www.rida.at/Wer-entwickelt-RIDA.321.0.html}. In the Netherlands there is Legal Intelligence owned by publisher Wolters Kluwer (\url{https://www.wolterskluwer.nl/shop/serie/legal-intelligence/Legal-Intelligence/}), and Rechtsorde owned by publisher Sdu (\url{https://www.sdu.nl/juridisch/producten-diensten/rechtsorde}), who in turn is part of publishing company Lefebvre Sarrut (\url{https://www.lefebvre-sarrut.eu/en/by-your-side/}). In Germany, there is Juris, owned in part by the German state and in part by Sdu (\url{https://www.juris.de/jportal/nav/juris\_2015/unternehmen\_2/ueber\_juris/ueber\_juris.jsp}) and thus by Lefebvre Sarrut, and Beck Online owned by C.H. Beck publishers (\url{https://beck-online.beck.de}).}
These commercial legal IR systems usually deal with multiple documents types. Systems like Westlaw, LexisNexis, and the Legal Intelligence system that we work with in this research, include not only laws and case law, but also legal journals, books, government reports and newspaper articles. 

\subsection{The Documents}\label{Documents}
When looking at legal IR systems with diverse document types, the large deviation in length of the documents in the index is often the most notable feature. Lengths may vary between a government report (161 pages)\footnote{DocumentID 34474736.} and a newspaper article (57 words)\footnote{DocumentID 34582268.}.\footnote{Note that books are often indexed by chapter or paragraph.} There is also a difference in genre, varying from the structured form of legal codes and case law, to the free form of blog posts and newspaper articles.

\par The scope of the collection of a legal IR system is smaller than in Web search, and pre-determined by the owner of the IR system. As mentioned above the collection is often limited to one legal jurisdiction. Documents included in the collection of a legal IR system are all from sources that are considered to be relevant to legal professionals. This restricted scope reduces noise, especially when dealing with homonyms. %The word `trust' for example in a legal context has a specific meaning: `The right, enforceable solely in equity, to the beneficial enjoyment of property to which another person holds the legal title; a property interest held by one person (the trustee) at the request of another (the settlor) for the benefit of a third party (the beneficiary).' \cite{Blacks}. 
To distinguish between the meaning of terms in ordinary speech and `legalese', law dictionaries are created, the most famous being Black's Law Dictionary \cite{Blacks}. By reducing the scope of the collection of the legal IR system to documents relevant to legal professionals, a search for `trust' by a legal professional will result in documents regarding this topic, rather than results about the company Trust and the character quality one might find in Web search\footnote{Incognito Google search conducted on October 30th 2020.}.%, matching the language use in queries to those in the collection.

\par A further narrowing of the scope of the collection comes from journals/sources with a subscription model. Where the government or a university is likely to purchase a blanket subscription to journals from all law areas, a niche law firm will likely subscribe to a limited amount of journals relevant to their work to limit expenses.\footnote{Though this depends on the price models used by the publishers, who sometimes price packages of content in such a way that a package deal with more content is cheaper than subscribing to only the journals needed.} Because of the difference in amount of documents accessible for each user, the same query will generate a different set of results for the lawyer than for the scholar.

\par When looking at the structure of the documents, it is noticeable that the reliance on legal codes and previous cases for argumentation means that there are a lot of references in legal documents. Though legal professionals have multiple methods to cite a document (e.g. party names, case number, journal reprint reference number), the various references can be mapped using regular expressions to provide an overview of the relations between documents. It appears though, that this information is not always used to the fullest extent possible \cite{Geist}. This in contrast to websearch, where PageRank has become the standard \cite{Manning}.

\subsection{Relevance} \label{Relevance}
IR, including legal IR, has as aim to aid users to find relevant information. For legal IR, this notion of relevance can be described by the following relevance factors, as identified in prior work \cite{Wiggers1}: title relevance, document type, recency \cite{Russel-Rose}, level of depth, legal hierarchy, law area (topic), authority (credibility), bibliographical relevance, source authority, usability, whether the document is annotated, and the length of the document. These relevance factors are similar to those in other fields, as demonstrated by the work of Barry and Schamber \cite{Barry1,Barry2}. Van Opijnen and Santos \cite{Opijnen} established that legal professionals tend to agree strongly on factors like authority, legal hierarchy and whether the document is annotated. While these factors are usually grouped under 'cognitive' or 'situational relevance' and thereby considered to be specific to the user or task, because of the general agreement between users in the legal domain on these factors, Van Opijnen en Santos \cite{Opijnen} group these as 'domain relevance'.

\subsection{Small Data} \label{Small Data}
Because of the time pressure users are under, and the associated labor costs, as mentioned in section \ref{Users}, it is often not possible for developers of legal IR systems to obtain large quantities of explicit feedback or relevance judgments. However, the use of implicit feedback collected in the course of normal search activities \cite{Kelly} is also limited, because legal IR systems are often bound to a particular jurisdiction. This means that the number of users in a system is limited to the legal professionals within that country. In the case of the Netherlands, the largest legal IR system has between 75 000 and 100 000 users. The amount of usage data available is therefor much lower than in IR systems for generic Web search.

\par This smaller dataset due to the size of the audience is narrowed even further when we remember that legal IR systems are not used daily. When we add to this a high attention to recency, as well as the differences in subscriptions, few users have seen the same results lists or query-results pairs. This means the data available for implicit feedback analysis is also limited.

\subsection{Legal Search and Professional Search}
Legal IR is a form of professional search, and shares many characteristics with it, as well as with other types of domain-specific search. The First International Workshop on Professional Search\footnote{Held at SIGIR 2018.} describes professional search as: ``professional search takes place in the work context, by specialists, and using specialist sources, often with controlled vocabularies." \cite{Verberne} It covers people from multiple domains, including librarians, scientists, lawyers, and other knowledge worker professions. They describe six characteristics: (1) a restricted scope and domain. Users do not wish to retrieve information from all possible sources, but only from within their domain (e.g. legal, medical). (2) Not all sources are equally accessible; subscriptions are required to access some sources. This means that two professionals with different subscriptions will retrieve different result sets. (3) the use of multiple systems; (4) a tolerance for low precision; professionals create lengthy queries and often take time to refine them. (5) the need for users to be in control: ``explaining the predominance of Boolean search in, e.g., prior art search and systematic review." \cite{Verberne} (6) the use of controlled vocabularies.

\par When applying these six characteristics to legal IR, we notice that (1) the restricted scope and (2) subscription access are indeed characteristics of legal IR, as shown in Section \ref{Documents}. Characteristic (3), the use of multiple systems, may vary from jurisdiction to jurisdiction. In countries like the United States and the Netherlands systems like Westlaw, LexisNexis and Legal Intelligence provide content integration as well as IR functionalities. Geist \cite{Geist} however describes that in Austria licensing issues have caused situations where legal IR systems include summaries of publications from other publishers in their index, but users must use the print version or change IR systems to be able to access the full-text of these documents.

\par As described in Section \ref{Users}, the (4) tolerance to low precision is described by Manning et al. \cite[p. 156]{Manning} to include legal IR, but debated by Geist \cite{Geist}. This is often related to (5) the need for control. Two well-known high recall tasks, often conducted using boolean queries for reproducability, are systemic review tasks (academic\footnote{For the purpose of this paper we will consider the search for scholarly information -- academic search -- part of professional search.}/medical search) and prior art search (patent search). However, several professional search domains, such as medical search and legal search, include instances of these high recall tasks, next to more applied search behaviours. The legal domain for example has a citation culture where legal scholarly articles may cite publications from legal practice \cite{Wiggers2}. The last characteristic, (6) the use of controlled vocabulary, is demonstrated by the existence of law dictionaries and has been discussed in Section \ref{Documents}.

\subsection{Summary}
\par Legal IR has several characteristics that challenge common evaluation methods: (1) The cost of missing results is high, but the tolerance to low precision results drops under time pressure. This means that early-precision metrics are not sufficient; lower-ranked documents also have to be part of the evaluation. (2) Explicit relevance judgements are expensive to gather. (3) Because the field of legal research is highly specific, the user group and number of user interactions is limited. (4) Different users see different results in their results list, based on the journals/sources they are subscribed to, and thus have access to. This limits the use of implicit feedback models further. (5) Recency is considered very important, and plays a large role in the ranking algorithm. Because of this, and the high frequency with which new documents are published, the top of the results list is highly dynamic.%, meaning that static evaluation methods are difficult to use for live systems.

\section{Expected challenges to common evaluation methods}
\label{Expected challenges}
All IR systems share the same aim: user satisfaction \cite{Manning}. This comprises multiple components, including speed, user interface\footnote{For the importance of snippets in Legal IR, see Wiggers et al. \cite{Wiggers1}.}, and satisfaction with the results returned. The satisfaction with the results returned depends on the number of relevant results returned, and the order in which the results are returned.
\par This research focuses on evaluation methods comparing two different versions of a ranking algorithm, in particular the following four common methods: (a) a test collection based on information needs and relevance judgments by domain experts, (b) a test collection based on implicit feedback, (c) user surveys, and (d) A/B testing. In the following subsections we discuss each of these in relation to our example: the evaluation of a live legal search engine.

\subsection{Test Collections}\label{Test Collections}
A common method of evaluation is test collections \cite{Jarvelin2009}. An example is the English language test collection for case law retrieval created by Locke and Zuccon \cite{Locke}, and the German language test collection created by Arora et al. \cite{Arora2018}. An initiative for benchmarking in legal IR is the Competition on Legal Information Extraction/Entailment (COLIEE), active since 2014 \cite{Rabelo}.\footnote{\url{https://sites.ualberta.ca/~rabelo/COLIEE2021/}} COLIEE's specific focus is on case law, using Canadian test collections. 

\par Conducting evaluations on these public test collections is less informative for legal search systems that cover non-English language civil law jurisdictions, as the content in the actual collection will be in the language of the jurisdiction, and the focus of the user may be more on legal statutes and less on case law. The evaluation of such a system on an English language test collection with a limited task (e.g. retrieving only case law or e-discovery) will provide little information on the performance of the system when used in daily legal practice in the home jurisdiction. In addition, case retrieval tasks such as the one in COLIEE are document-to-document tasks, where the query is a case law document, as opposed to a keyword query. Most commercial professional search engines, including ours, use keyword queries.

\par Hawking \cite{Hawking} suggests that a test collection for professional search (in his situation enterprise search) should be created specifically for the company in order to be a suitable evaluation method. The set will have to be tailored to the company because of the highly specialized content used in the system.

\par Conrad and Zeleznikow \cite{Conrad} mention that relevance assessments are often created by some sort of domain expert, for example grad students or research assistants. However, as Cole and Kuhlthau \cite{Cole} have shown, there is a difference between what an early career legal professional classifies as relevant, and what a senior legal professional classifies as relevant, in line with the notion of cognitive relevance of Saracevic \cite{Saracevic}. This is also the reason why relevance assessments are usually gathered from multiple assessors. In the case of legal professionals, that would require relevance judgments of not only junior but also (more expensive) senior legal professionals, as well as participation from scholars and the judiciary.

\par As stated by Voorhees \cite{Voorhees2}, for many evaluation metrics used in test collections, a substantial number of documents in the results list need to be judged. Even when using %within-system 
pooling, as in the case of Arora et al. \cite{Arora2018} for professional information needs, these judgements need to be made by multiple assessors. Since this requires the inclusion of high level experts as described above, this becomes prohibitively expensive. In addition, since we are dealing with information needs in the legal domain that potentially require a high recall, we need deep relevance assessments and a shallow pool with only a few assessed documents per query does not suffice.

\par An alternative to using test collections with expert judgments is the use of implicit feedback. In Section~\ref{Found challenges} we will assess the value of test collections based on explicit or implicit feedback for the evaluation of a live professional search engine.

\subsection{User surveys}\label{Survey}
Asking a user directly whether they are satisfied provides valuable information. However, the research of Blair and Maron \cite{Blair} suggests that there is likely to be a mismatch between the recall the users think they have achieved and the recall calculated based on random samples of documents in the collection. In their research with legal professionals the average calculated recall was 20 percent, whereas the legal professionals questioned believed they were at 75 percent recall or higher.

Furthermore, as suggested by Turpin and Hersh \cite{Turpin}, a ranking that scores higher on system oriented metrics does not always score higher using user oriented evaluation metrics \cite{zhang2020}. Literature suggests this to be especially true when the difference between the rankings is small and not at the extreme ends of performance (e.g. both are not extremely poor systems or extremely good systems) \cite{Smith}. Users can adapt their search strategies to achieve similar levels of results for different levels of quality systems \cite{Allan}, for example by refining their queries \cite{Turpin} This might be a limitation for use as an evaluation method for professional search systems, as a commercial system is unlikely to be an extremely poor system, and a change to the ranking algorithm is unlikely to create drastic changes such as a complete reversal of the ranking.

\par For commercial websites and webservices, measuring user satisfaction is often done through Reichheld's Net Promotor Score \cite{Reichheld}, a very short survey that measures user satisfaction. The appeal of the Net Promotor Score (NPS) as compared to other types of surveys is that the shortness makes for a higher response rate. 
\par It should be noted that Reichheld shows that the NPS score has a lower correlation with sales where the purchase decision is not made by the individual user, but by company management, such as computer systems \cite[p. 6]{Reichheld}. It is therefore important to carefully consider the framing of the question in a manner that corresponds with the information desired.

\par In Section~\ref{Found challenges} we fill assess the value of two types of user surveys -- a ranking preferences survey and an NPS survey -- for our example.

\subsection{A/B testing}\label{sec:ABtesting}
For large scale systems like Google, the evaluation is often done with live user-oriented evaluation methods in the form of an A/B test \cite{Tang}. A/B testing is a between-group design that usually consists of (1) randomly splitting the users into two representative groups, a test group and a control group, and (2) presenting the test group a feature (whether in the interface or in the ranking algorithm) while keeping the control group on the current version of the system \cite{Tang}. The two groups are then compared on variables such as user engagement.

\par The legal domain has both users that search for themselves and users (e.g. paralegals) that search for others. In conversations with management of the Legal Intelligence system we found that customers expect the system to return the same results for all users. This so that the work of the paralegal or intern can be replicated and checked. Therefore, in the legal domain, it is commercially not acceptable to differentiate between users from the same organization. When trying to split the user group on organizational level, we found that due to the many firms who specialize in one area of the law, it is difficult to create two groups that are both representative. There is also commercial pressure to provide the latest (and thereby believed to be best) version of the system to all customers. For these commercial reasons it is not possible to divide the entire customer base of a live system into two groups, whether on user or on organisation level. This appears to be a blocking factor for using A/B testing in practice. 
\vspace{0.5cm}

\par This means that we have three evaluation methods left (test collections based on expert judgments, test collections based on implicit feedback, and user surveys), which we will apply and assess for our example based on data from the search engine and user studies.

\section{Assessment of applied evaluation methods}\label{Found challenges}
In this section we show, supported by descriptive statistics of data from the search engine and two user studies, the implications of applying common evaluation methods to a live professional search engine: (a) a test collection based on expert relevance judgments (Section~\ref{sec:expert}), (b) a test collection based on implicit feedback (Section~\ref{sec:ImplicitFeedback}), and (c) two surveys: a survey measuring users' preferences for rankings (Section~\ref{sec:survey}), and a survey based on the Net Promotor Score (Section~\ref{sec:NPS}). For each method we discuss the suitability and challenges of the method for legal IR in practice, with a focus on monitoring the effectiveness of changes to a single legal IR system over time.

\subsection{Test collection based on expert relevance judgments}\label{sec:expert}
In the case of Legal Intelligence, an early precision (or shallow pool) golden standard, or golden data set, internally known as the `golden answer set', is available. This data set contains queries and their `golden answers'; documents that are expected to be the top ranked results. This set of queries and their corresponding golden answers has been created by editors of legal journals, who are domain experts in their law area. The set contains 194 queries with for each query between 1 and 17 golden answers. The collection has been build by sampling from queries conducted by domain experts in the past, eliciting the results they would have liked to have seen in top positions. This set is subdivided into case law (51 queries), literature (51 queries), legal codes (46 queries) and legal commentary (46 queries).
\par Because this data set focuses on early precision through golden answers (results expected on top positions), it does not contain relevance judgments for all results returned. This requires less relevance judgments, and is therefore cheaper to make. This is, however, also the most important limitation of this method. Because the set is only limited to only a small number of relevance judgments, this tool cannot be used to assess the ranking algorithm for high recall scenarios. The use of this set is limited to `research reality' scenarios as described by Geist \cite{Geist} where the focus is on early precision.
\par Further challenges include the age of the set. The set was created in 2018, meaning that newer, perhaps more relevant results, have not been included. Regularly updating this data set is time intensive, and therefore expensive. In practice, the problems with the age of the judgements are circumvented by using a document collection with publication dates up until 2018, and pretending it is early 2019 to ensure that date boosts are functioning correctly. While this method allows developers an easy way to compare two versions of a ranking, this clearly does not reflect the reality that the top of the results list is highly dynamic. This limitation exists for all test collections, but is more prominent when using the method for the evaluation of continuing updates to a single system.

\par An early precision golden data set does not provide information that can be used to infer pairwise preferences: document A is expected above B, but when B is also marked as relevant, that cannot be taken to mean that either A or B in isolation does not provide sufficient information for the information need behind this query, as that was not considered when creating this test collection. A further limitation is the subscription model used for legal publications. The document marked most relevant for the query may be outside the subscription of the user. If no alternative document has been marked as `second best', the golden standard set does not reflect the user experience of users who are not subscribed to the publication this document appeared in.
\par Because of these challenges, the golden standard set is only suitable for developers to conduct sanity checks when developing a new ranking algorithm, taking into account that the results only reflect early precision use cases, not high recall use cases. An updated test collection with relevance assessments done by multiply users including senior legal professionals is too expensive. Test collections are therefore not a viable method to evaluate changes made to the ranking algorithm of legal IR systems beyond technical sanity checks.

\subsection{Test collection based on implicit feedback}\label{sec:ImplicitFeedback}
Implicit feedback appears promising because, unlike the test collection mentioned above, it does not require a time investment from the users or domain experts and is usually readily available in legal IR systems. As it is collected during the normal work process of the user, the data is always up-to-date.

\par Implicit relevance judgments can be used to infer relevance from (user) interactions. In the Netherlands Van Opijnen \cite{Opijnen2} studied implicit feedback as signal for the relevance of case law. This work focused mainly on (re)publication as signal rather than user interaction.

\par Addressing the interactions of users with the search engine, Oard and Kim \cite{Oard2001} have created a framework that describes the different types of user behaviour that could be monitored for implicit feedback. Methods that have been proposed to assign relevance scores to documents include Click Through Rate (CTR) and pairwise inference (see e.g. Joachims et al. \cite{Joachims2}, further expanded on by Chuklin et al. \cite{Chuklin} and Agrawal et al. \cite{Agrawal}).

\par The implicit feedback data that we use contains the clicks registered in the logs of the Legal Intelligence system, with a pseudonomized user ID, the document ID, the position of the document in the ranking, the text of the query and a datestamp.
\par The search engine result page of Legal Intelligence contains links to 20 documents. When a user scrolls to the bottom of the results page, a further 20 results will be loaded, if available. Each document is described by a publisher curated abstract that consists of the title of the document and varying amounts of meta-data. When a user clicks on a result, they will be directed towards the full article on the platform of the publisher of the article. Because the user is outside the Legal Intelligence system while reading the article, and is able to click through to other articles while on the publisher platform, reading time is not logged in the Legal Intelligence logs; we only use clicks as the signal of (implicit) relevance.

\begin{figure}[t]
\includegraphics[width=0.8\textwidth]{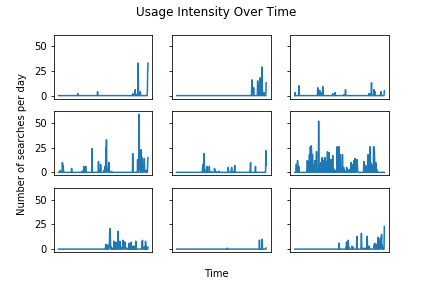}
\caption{The number of searches per day for nine users over 10 months.} \label{fig2}
\end{figure}

\paragraph{The amount of data per user}\label{Daily} To explore the data available to a commercial legal IR system, and the challenges it causes, we looked at the patterns of user interactions per user. To give an example of the activity of users of legal IR systems, and how much data they generate per person in their day to day activities, we selected the nine users who conducted the most recent queries reported in the logs. For these nine users, we tracked the number of queries in the Legal Intelligence system from the first of January 2020 to the 20th of October 2020. Figure \ref{fig2} shows the usage patterns of these nine users. Though the average number of queries varies between users, all users show periods of more intense research and periods of less intense research. This shows that of the total user group, only a part is active on a given day.

\paragraph{Queries are usually unique}\label{Multiple} We also looked at the number of queries that have been issued by multiple users within one month. We zoom in on a period of one month (October 2020), because of the highly dynamic top of the results list, as discussed in Section \ref{Relevance}.

\begin{figure}[t]
\begin{floatrow}
\ffigbox{%
  \includegraphics[width=1\linewidth]{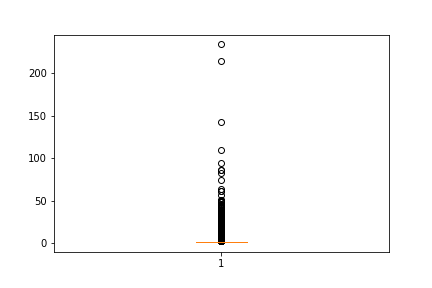}
}{%
  \caption{Distribution of queries and number of users conducting them}
  \label{Numbers}
}
\capbtabbox{%
  \begin{tabular}{|p{2.5cm}|r|} 
  \hline
 & No. of users issuing  \\ 
 & the same query \\ 

 \hline
mean & 1.16\\
std & 1.25\\
min & 1.00\\
25\% & 1.00\\
50\% & 1.00\\
75\% & 1.00\\
max	& 234\\
\hline
  \end{tabular}
}{%
  \caption{Distribution of queries and number of users conducting them}
  \label{TableNumbers}
}
\end{floatrow}
\end{figure}

\par To create implicit feedback models, whether through click-through rates or pairwise inferences, we generally need queries that are conducted by multiple users. We need enough data to rate the entire results list, or the @k results specified in the evaluation metric. But we also need enough data to compensate for the fact that the users may have seen different results list due to differences in subscription or new documents being added to the collection. Different result lists mean users have seen different pairs of results and generate different pairwise inferences. As shown in Figure \ref{Numbers} and Table \ref{TableNumbers} the majority of queries is unique to one user. This is not unexpected as professional search deals with experts. It is not unreasonable to assume that the more expertise a user has on a topic, the more unique the queries become \cite{Cole,Verberne}.

\paragraph{Queries issued by multiple users}\label{Generic} When we look at the top 10 queries ordered by number of users that conducted that query, we see in Table \ref{TablePopularQueries} that these queries are often navigational queries where a user wishes to find and open a particular source, for example a book or journal. We consider this separately from known item retrieval, where the user wants to access a particular document from that source, for example an article or chapter. In Table \ref{TablePopularQueries} this difference is illustrated by queries on source names `lexicon', `tekst en commentaar', `asser' and `wpnr', and queries for sepcific documents `awb' and `ECLI:NL:HR:2013:BZ2653'. These navigational queries would provide a very one-sided image of legal IR if used in implicit feedback models.

\begin{table}[t]
\centering
 \begin{tabular}{|p{8cm}|p{2.5cm}|} 
  \hline
 Query & Number of Users \\ 
 \hline
poging tot doodslag (`attempted homicide') & 234 \\
$^{*}$ (a wildcard query to retrieve all documents)\footnotemark & 215\\
lexicon (source name) & 142\\
tekst en commentaar (source name) & 109\\
onrechtmatige daad (`tort') & 94\\
awb (law name)& 86\\
corona (colloquial reference to the SARS-COV-2 virus) & 86\\
asser (source name) & 83\\
ECLI:NL:HR:2013:BZ2653 (case law identifier)& 74\\
wpnr (source name) & 64\\
\hline
  \end{tabular}
  \caption{Distribution of Queries and Number of Users Conducting Them}
  \label{TablePopularQueries}
\end{table}
\footnotetext{Users may use this if they wish to navigate using filters rather than a query.}

%\par These results show that our case study is indeed confronted with small data in the context of evaluation methods and metrics, which makes a test collection with a full ranked results list based on implicit feedback impossible due to a lack of data. To summarize, this is caused by several factors:
%\begin{enumerate}
    %\item The amount of users, and their usage of the system, is limited.
    %\item The amount of queries done by multiple users is small because of the specific nature of expert search, meaning the number of queries with enough click data to use for implicit feedback assessments is small. 
    %\item The queries that are issued by multiple users are often navigational or not very specified, meaning an evaluation on these queries is not representative for the total user experience. 
    %\item Due to differences in subscription, users did not see the same results for the same query. This means users do not see the same pairs of results and makes it very hard to infer relevance judgments through pairwise preferences.
    %\item The top of the results lists changes often (see Section \ref{Relevance}), so only recent queries and clicks can be used in implicit feedback, as the top 20 seen today is not the same as the top 20 seen last month. 
%\end{enumerate}
\par This means that using implicit feedback to infer relevance for test collections, even in the case of partially judged results lists, is not viable for a professional search system like Legal Intelligence.

\subsection{Survey for ranking preferences}\label{sec:survey}
To asses surveys as an evaluation method, we created one. The survey was made using compilations of screenshots from the search engine. It shows the query, followed by two images of result lists, as shown in Figure \ref{figSurvey} in Appendix A. Respondents are asked to indicate which ranking they prefer. Respondents also have the option to indicate that they have no preference.

\par The two rankings used are a baseline ranking (the then current ranking in the legal IR system) and a degraded model, inspired by Smith and Kantor \cite{Smith}. In our test set up the degraded model was created by removing boost functions from the baseline model of which we know that they were added at the request of users. Thus, we know that the degraded model differs in a manner relevant to the user. We chose a relative relevance assessment method (``which of the two rankings do you prefer''), since it has been demonstrated that humans can make such relative decisions more reliably \cite{Fox}, and it helps negate the bias of work experience \cite{Saracevic}. 

\paragraph{Survey design} As per TREC convention, we aimed at 50 reviewed query/rankings pairs (QRPs), with a minimum of 25 reviewed pairs \cite{Voorhees2} and a minimum of 3 respondents per pair. The QRPs were divided per law area. Users were asked to indicate the law area they practice in, and were shown QRP's accordingly. This was done to ensure experts in a particular legal domain reviewed only QRPs for which they were able to assess the information need behind the query, and the relevance of the results for the query. We also include general practice queries for which respondents were able to asses the general information needs. 

\par We selected queries that multiple users have issued, from multiple companies, to avoid privacy sensitive queries. This also reduces the risk of noise by accidental clicks. As shown in Section \ref{Generic} queries issues by multiple users tend to either be less specific or navigational. If those are used in an evaluation method they will give an incomplete image of the quality of the system, but in the context of testing whether users can agree on a preferred ranking the general nature of the queries may be helpful as it will allow users to understand the information need behind the query. We selected queries from 7 law areas: corporate law, IT law, environmental law, labour law, tax law, criminal law and generic legal practice. Each law area included at least one query for a law article, one query for a law name, and one or more queries for a legal concept. Each set of queries included one query (except the general group, which had two) that was also included in one of the other sets, leading to a total of 55 different queries. With these 55 queries we created 9 QRPs per law area for 7 law areas, for a total of 63 pairs.
 
\par Respondents were given 9 QRPs to review, each with two rankings of 10 results, but were able to end the survey earlier. We decided to allow this to ensure the largest number of participants possible.

\par We inspected the rankings to confirm that they are different. On average 2.4 documents in the top-10 remained in the same position, whilst 7.6 documents changed position. Of these 7.6 documents 1.4 documents moved up, 2.9 moved down, and 3.2 were replaced. However, as Table \ref{table:Surveychanges} in Appendix A shows, in some cases the results list of the degraded model had no documents in common with the results list of the baseline model. To show that the changes in the order of results were relevant, we created a highly simplified implicit feedback model. As shown in Section \ref{sec:ImplicitFeedback} we only had generic queries that were done by multiple persons, and for those we had on average a total of 3.7 clicks (from all users combined) in the top 20 to base our nDCG calculation on. We considered a clicked document to be relevant, and an un-clicked document to be neutral. Using this click data we calculated the nDCG@20 under the old and new ranking. This was 2.08 for the old ranking, and 1.96 for the new ranking. While we expected the nDCG to reflect that the degraded model, because we removed boosts added to the system at the request of the users, was less preferable, the score suggests otherwise. However, for the purpose of this survey the question is not which is better, but whether users see a difference, and indicate the same preferences.

\par The order of the baseline model and the degraded model was alternated. Our hypothesis is that if the survey is an appropriate evaluation tool, users will notice difference between the the two rankings and indicate a preference for one of the rankings.

\begin{table}[t]
\centering
\titlerunning{User Preference Compared to Baseline and Degraded Ranking (excluding no preference)}
\begin{tabular}{|l|l|l|}
\hline
Users Prefer Baseline & Users Prefer Degraded & Users Tie\\
\hline
29 & 23 & 11\\
\hline
\end{tabular}
\caption{Number of QRPs (total 63) by majority preference. Users considered tied when number of users indicating choice 1 and 2 is equal (regardless of number of no preferences)}
\label{table:Surveynotie}
\end{table}

\begin{table}[t]
\centering
\titlerunning{User Preference Compared to Baseline and Degraded Ranking (including no preference)}
\begin{tabular}{|l|l|l|}
\hline
Users Prefer Baseline & Users Prefer Degraded & Users Tie\\
\hline
12 & 9 & 42\\
\hline
\end{tabular}
\caption{Number of QRPs (total 63) by majority preference. Users considered tied when number of users indicating no preferences is higher than choice 1 and/or 2}
\label{table:SurveywithTie}
\end{table}

\paragraph{Results} The survey was completed by 77 respondents. Each of the 7 law areas had at least 3 respondents. For our analysis, we selected the majority answer for each of the 63 QRPs. In Table \ref{table:Surveynotie} we excluded the answers from respondents who indicated that they had no preference; in Table \ref{table:SurveywithTie} we considered the pair also tied when the number of respondents indicating no preference was higher than the number of users indicating option 1 or 2. 

\paragraph{Analysis} We expected to find a preference from the users for one ranking over the other, as the nDCG scores indicated that the relevant documents had moved, and the change we made to create the degraded model was a boost function introduced at the request of the users and as such is expected to be noticeable by the users. As shown in Table \ref{table:Surveynotie} and Table \ref{table:SurveywithTie}, this was not always the case. This means that a survey of this kind does not elicit enough information to base an evaluation on. We conclude that a ranking preference survey is not a usable evaluation method for our example.

\subsection{Survey based on the Net Promotor Score}\label{sec:NPS} As a second type of survey, we experimented with the Net Promoter Score (NPS) as described in Section \ref{Survey}. We chose this type of survey because of the low user effort required. The NPS data is constantly being collected for commercial purposes, meaning the data is readily available. The NPS measures overall user satisfaction, and does not focus specifically on the ranking. However, one would expect that an improvement in the ranking of the search results would also improve the overall user satisfaction.

\par For our experiment we chose a real live change in the ranking algorithm of the Legal Intelligence system that went live on September 14th 2020. In our situation the NPS score is gathered per month, so we compared August 2020 with October 2020. The NPS question is not always presented to users. To avoid irritating users the question is posed at the most once every six months. Furthermore a user has to be logged in to see the NPS question. As shown in Section \ref{Daily}, users do not use the system daily, so the user population that is shown the NPS question on a given day is small. Of the users that are shown the NPS question, not all respond. In both months, ten users responded to the NPS survey.

\par The scores were exactly the same for both months.\footnote{Because of commercial interests the exact NPS score cannot be reported in this paper.} Like with the other survey, this may be explained by the difference being small, and because of the adaptability of research strategies by users. However, the broadness of the measure and the low number of respondents suggests that the NPS is not a good approach to assess differences in ranking within a legal IR system, especially for jurisdictions of a modest size.

\section{Conclusion}\label{Conclusion}
Legal professionals are confronted with information overload, and are in need of effective legal IR systems. Though evaluation of these systems is considered important from an academic point of view, in practice this is not always conducted in a consistent manner. In this paper we showed, using data from a live professional search system, the challenges of common evaluation methods.

\par The focus of this research is on situations where a change is made to the algorithm that affects the ranking of the documents but not the number of documents retrieved or other changes to the IR system, including the user interface. Its application is therefore limited to within-system comparisons, not between--systems comparisons. The applicability of our work is limited to commercial, medium-sized professional IR systems.

\par The common evaluation methods were defined as: (a) a test collection based on information needs and relevance judgments by domain experts, (b) a test collection based on implicit feedback from clickthrough/log analysis, (c) user satisfaction studies (in particular surveys), and (d) A/B testing.

\par As argued in Section \ref{sec:ABtesting}, A/B testing is not an option because in the legal domain commercial reasons prohibit different users seeing different results. As shown in Section \ref{sec:expert} test collections based on relevance judgments from domain experts are too expensive to gather and keep up to date. Implicit feedback data is also not suitable for creating test collections, as the available data is too sparse, in particular with regards to queries issued by multiple users, as shown in Section \ref{sec:ImplicitFeedback}

\par As shown in Section \ref{sec:survey}, surveys are not a suitable evaluation method to evaluate differences in ranking algorithms in legal IR. The survey on ranking preferences in our legal search engine showed inconclusive results. The NPS survey analysis shows that the number of users exposed to the NPS questions and the broad nature of the question make it not suitable.

\par Given the found challenges, we find that all of the common evaluation methods are sub-optimal for use in evaluating changes to ranking algorithms in live professional information retrieval systems. In our future work we will focus on less common evaluation methods, such as a cost-based evaluation model as described by J\"{a}rvelin et al. \cite{Jarvelin2009}.

%
%--- Bibliography ----
%
% BibTeX users should specify bibliography style `splncs04'.

\appendix
\section{Appendix A}

This table shows the the difference between the baseline ranking and the degraded ranking. Each row in the table represents one query. Each ranking shown was 10 documents. Each number in the column corresponds with the number of documents for that query that was ranked higher or lower in the degraded model, documents that were not present in the degraded model but replaced by another document, and the number of documents that remained in the same position.

\begin{table}[]
\centering
\titlerunning{Difference between baseline and degraded ranking}
\begin{tabular}{|l|l|l|l|l|}
\hline
QueryID & Moved Up & Moved Down & Document Replaced & Same Position\\
\hline
1&1&2&2&5\\
2&3&4&1&2\\
3&1&6&2&1\\
4&1&4&3&2\\
5&0&4&6&0\\
6&0&2&8&0\\
7&0&2&6&2\\
8&2&2&1&5\\
9&3&4&1&2\\
10&2&3&3&2\\
11&2&3&4&1\\
12&4&2&2&2\\
13&2&3&4&1\\
14&1&1&1&7\\
15&1&4&4&1\\
16&1&2&5&2\\
17&0&5&4&1\\
18&1&4&3&2\\
19&1&3&1&5\\
20&3&3&1&3\\
21&1&3&2&4\\
22&3&3&1&3\\
23&4&3&1&2\\
24&0&1&4&5\\
25&1&1&2&6\\
26&1&2&3&4\\
27&0&7&2&1\\
28&2&1&2&5\\
29&1&4&3&2\\
30&3&2&3&2\\
31&3&2&3&2\\
32&0&3&3&4\\
33&3&4&3&0\\
34&1&4&2&3\\
35&2&3&3&2\\
36&0&7&3&0\\
37&2&0&2&6\\
38&0&0&10&0\\
39&2&1&3&4\\
40&1&6&3&0\\
41&2&3&4&1\\
42&0&6&4&0\\
43&0&6&2&2\\
44&0&2&6&2\\
45&0&0&10&0\\
46&1&0&9&0\\
47&3&3&2&2\\
48&3&0&2&5\\
49&0&5&3&2\\
50&2&1&1&6\\
51&0&2&2&6\\
52&5&1&3&1\\
53&3&2&2&3\\
54&1&3&5&1\\
55&2&5&2&1\\
\hline
\end{tabular}
\caption{The number of the documents from the top-10 that changed position in the degraded ranking as compared to the baseline ranking.}
\label{table:Surveychanges}
\end{table}

\begin{figure}
\includegraphics[width=\textwidth]{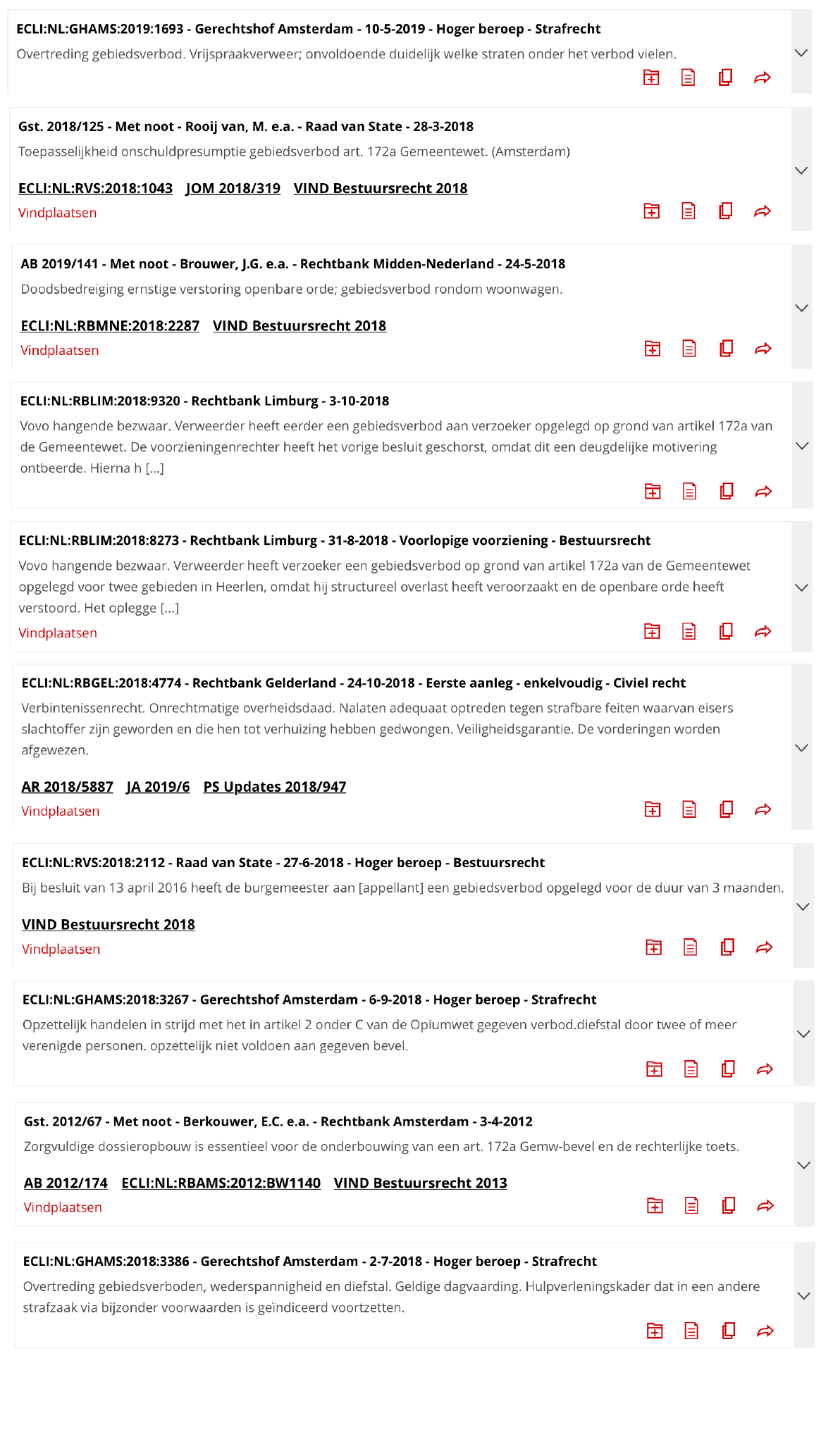}
\caption{An example of a result list as displayed in the survey.} \label{figSurvey}
\end{figure}

\end{document}